\input harvmac
\overfullrule=0pt
\abovedisplayskip=12pt plus 3pt minus 3pt
\belowdisplayskip=12pt plus 3pt minus 3pt
%

\def\to{\rightarrow}
\font\zfont = cmss10 
\font\litfont = cmr6

\def\bigone{\hbox{1\kern -.23em {\rm l}}}
\def\ZZ{\hbox{\zfont Z\kern-.4emZ}}
\def\half{{\litfont {1 \over 2}}}

\def\mg1{{\cal M}_{g,1}}

\def\cmg1{{\overline{\cal M}}_{g,1}}


\def\BST{E. Bergshoeff, E. Sezgin, and P.K. Townsend, Phys. Lett. 
{\bf B189} (1987) 75.}
\def\DHIS{M.J. Duff, P.S. Howe, T. Inami, and K.S. Stelle,
Phys. Lett. {\bf B191} (1987) 70.}
\def\TOWNA{P.K. Townsend, hep-th/9501068, 
Phys. Lett. {\bf B350} (1995) 184.}
\def\WITTENA{E. Witten, hep-th/9503124, Nucl. Phys. {\bf B443} 
(1995) 85.}
\def\BARS{I. Bars, hep-th/9503228, Phys.Rev. {\bf D52} (1995) 3567.}
\def\HARSTR{J.A. Harvey and A. Strominger, {\it The Heterotic String
is a Soliton}, hep-th/9504047.}
\def\TOWNB{P.K. Townsend, hep-th/9504095, Phys. Lett. {\bf B354} 
(1995) 247.}
\def\HULLTOWN{C. Hull and P. Townsend, hep-th/9410167, Nucl. Phys. 
{\bf B438} (1995) 109.}
\def\DUFFLIU{M.J. Duff, J.T. Liu, and R. Minasian, {\it Eleven
Dimensional Origin of String/String Duality: a One Loop
Test}, hep-th/9506126.}
\def\TOWNC{P.K. Townsend, {\it P-Brane Democracy}, hep-th/9507048.}
\def\BECKER{K. Becker, M. Becker, and A. Strominger,
{\it Fivebranes, Membranes, and Non-perturbative String Theory},
hep-th/9507158.}
\def\ASPINA{P. Aspinwall, {\it Some Relationships Between
Dualities in String Theory}, hep-th/9508154.}
\def\SCHWARZA{J.H. Schwarz, {\it An SL(2,Z) Multiplet of Type II
Superstrings},\hfill\break hep-th/9508143\semi
J.H. Schwarz, {\it Superstring Dualities}, hep-th/9509148\semi
J.H. Schwarz, {\it The Power of M Theory}, hep-th/9510086.}
\def\HORWIT{P. Horava and E. Witten, {\it Heterotic and Type I String 
Dynamics From Eleven Dimensions}, hep-th/9510209.}
\def\BARYANK{I. Bars and S. Yankielowicz, {\it U-Duality Multiplets and 
Nonperturbative Superstring States}, hep-th/9511098.}
\def\DLPS{M.J. Duff, H. Lu, C.N. Pope and E. Sezgin, {\it Supermembranes 
With Fewer Supersymmetries}, hep-th/9511162.}
\def\MAHA{J. Maharana, {\it M Theory and p-branes}, hep-th/9511159.}
\def\KMP{S. Kar, J. Maharana and S. Panda, {\it Dualities in 
Five Dimensions and Charged String Solutions}, hep-th/9511213.}
\def\TOWND{P.K. Townsend, {\it D-branes from M-branes}, hep-th/9512062.}

\def\DNP{M.J. Duff, B.E.W. Nilsson and C. Pope, Phys. Lett {\bf B129}
(1983) 39\semi
M.J. Duff and B.E.W. Nilsson, Phys. Lett {\bf B175} (1986) 417.}
\def\DUFFCY{M.J. Duff, in {\it Architecture of Fundamental
Interactions at Short Distances}, Les Houches Lectures (1985), 
eds. P. Ramond and R. Stora (North Holland).}
\def\HOWETOWN{P.S. Howe and P.K. Townsend, {\it Supermembranes and the 
Modulus Space of D = 4 Superstrings}, in {\it Supermembranes and
Physics in 2+1 Dimensions}, eds. M.J. Duff, C.N. Pope and E. Sezgin
(World Scientific 1990).} 
\def\CCDF{A.C. Cadavid, A. Ceresole, R. D'Auria and S. Ferrara, 
hep-th/9506144, Phys.Lett. {\bf B357} (1995) 76.}
\def\PAPTOWN{G. Papadopoulos and P.K. Townsend, hep-th/9506150, 
Phys.Lett. {\bf B357} (1995) 300.}
\def\HORAVA{P. Horava, Nucl. Phys. {\bf B327} (1989) 461.}
\def\HORETC{A. Sagnotti, {\it Open Strings and Their Symmetry Groups},
in {\it Non-perturbative Quantum Field Theory}, Cargese 1987, eds. G.
Mack et. al. (Pergamon Press 1988)\semi
P. Horava, Phys. Lett. {\bf B231} (1989) 251\semi
J. Dai, R.G. Leigh and J. Polchinski, Mod. Phys. Lett. {\bf A4} (1989)
2073.}
\def\TOWNE{P.K. Townsend, Phys. Lett. {\bf B139} (1984) 283.}
\def\SCHWHOPK{J. Schwarz, {\it Gauge Groups for Type I Superstrings},
in {\it Lattice Gauge Theories, Supersymmetry and Grand Unification}, 
Proc. Johns Hopkins Workshop, Florence 1982.}
\def\POLCOMB{J. Polchinski, hep-th/9407031, Phys. Rev. {\bf D50} 
(1994) 6041.}
\def\POLWIT{J. Polchinski and E. Witten, {\it Evidence for
Heterotic-Type I String Duality}, hep-th/9510169.}
\def\SEIBERG{N. Seiberg, Nucl. Phys. {\bf B303} (1988) 286.}
\def\ASPMOR{P.S. Aspinwall and D.R. Morrison, {\it String Theory on
K3 Surfaces}, hep-th/9404151.}
\def\DABHOLKAR{A. Dabholkar, hep-th/9506160, Phys.Lett. {\bf B357}
(1995) 307.}
\def\HULL{C.M. Hull, hep-th/9506194, Phys.Lett. {\bf B357} (1995) 545.}
\def\ALVWIT{L. Alvarez-Gaum\'e and E. Witten, Nucl. Phys. {\bf B234}
(1983) 269.}
\def\NARAIN{K.S. Narain, Phys. Lett. {\bf B169} (1986) 41.} 
\def\MARCSAG{N. Marcus and A. Sagnotti, Phys. Lett. {\bf B188} (1987)
58.} 
\def\WITNEW{E. Witten, {\it Five-branes and M-Theory on an Orbifold},
hep-th/9512219.}

{\nopagenumbers
\Title{\vtop{\hbox{hep-th/9512196}
\hbox{TIFR/TH/95-64}}}
{\centerline{Orbifolds of M-theory}}
\centerline{Keshav Dasgupta\foot{E-mail: keshav@theory.tifr.res.in}
and Sunil Mukhi\foot{E-mail: mukhi@theory.tifr.res.in}}
\vskip 4pt
\centerline{\it Tata Institute of Fundamental Research,}
\centerline{\it Homi Bhabha Rd, Bombay 400 005, 
India}
\ \medskip
\centerline{ABSTRACT}

We study $Z_2$-orbifolds of 11-dimensional M-theory on tori of various
dimensions. The most interesting model (besides the known $S^1/Z_2$
case) corresponds to $T^5/Z_2$, for which we argue that the
resulting six-dimensional theory is equivalent to the type IIB
string compactified on K3. Gravitational anomaly cancellation plays
a crucial role in determining what states appear in the twisted
sector. Most of the other models appear to break spacetime
supersymmetry. We observe that M-theory tends to produce chiral
compactifications on orbifolds, and that our results may provide an
insight into the mechanism by which twisted-sector states arise in
this hypothetical theory.

\ \vfill 
\leftline{December 1995}
\eject} 
\ftno=0
\newsec{Introduction} 
An 11-dimensional supersymmetric theory of gravity appears to exist
from which many (perhaps all) properties of string theory can be
extracted[1-19]%
\nref\bst{\BST}\nref\dhis{\DHIS}\nref\towna{\TOWNA}%
\nref\wittena{\WITTENA}\nref\bars{\BARS}\nref\harstr{\HARSTR}%
\nref\townb{\TOWNB}\nref\hulltown{\HULLTOWN}%
\nref\duffliu{\DUFFLIU}\nref\townc{\TOWNC}%
\nref\becker{\BECKER}\nref\aspina{\ASPINA}\nref\schwarza{\SCHWARZA}%
\nref\horwit{\HORWIT}\nref\baryank{\BARYANK}\nref\dlps{\DLPS}%
\nref\maha{\MAHA}\nref\kmp{\KMP}\nref\townd{\TOWND}. 
This theory (currently called ``M-theory'') has 11-dimensional
supergravity as its low energy effective field theory,
and does not appear to be described by strings. Certain properties of
M-theory can be described in terms of extended objects, the twobranes
and fivebranes, much in the same way that properties of various string
theories can be described by $p$-branes for various $p$.

Because M-theory is not a string theory, most of its compactifications
that have been studied are geometric in nature, since the analogue of
conformal field theory (if any) relevant to this case is not known.
These include compactifications on tori,
K3\ref\dnp{\DNP}\wittena\townb, Calabi-Yau
spaces\ref\duffcy{\DUFFCY}\ref\howetown{\HOWETOWN}\ref\ccdf{\CCDF},
and other manifolds of exceptional holonomy\ref\paptown{\PAPTOWN}.
One notable exception appeared in recent work of Horava and
Witten\horwit, where a proposal was given for the compactification of
M-theory to 10 dimensions on the orbifold $S^1/Z_2$. This
construction, while closely following an earlier construction of
``orientifolds''\ref\horava{\HORAVA}\ref\horetc{\HORETC}\ in closed
string theories, is necessarily more difficult to study since the
properties of M-theory responsible for non-geometrical effects are
presently unknown. Nevertheless, a number of consistency conditions
enabled the authors of Ref.\horwit\ to conclude that there should be a
sensible orientifold of M-theory on $S^1/Z_2$, leading to the
$E_8\times E_8$ heterotic string in 10 dimensions.

In the following, we will explore other ways of orbifolding and
orientifolding M-theory, with a view to learning what kind of
non-geometric behaviour of the underlying theory can be inferred.
Consistency conditions will as usual be crucial in drawing conclusions
for which calculational methods are as yet unavailable; these
conclusions will of course be worth testing should such methods become
available in the future.

It is appropriate to briefly review the orbifold/orientifold
ideas\horava\horetc\ which played an important role in
Ref.\horwit. The basic observation is that an orientation-reversing
$Z_2$ transformation of oriented closed string world sheets is a
symmetry in 10-dimensional type IIB string theory, and can be realized
also on type IIA string theory if additionally one spacetime dimension
is compactified and changes sign under the $Z_2$ action. In the former
case one discovers a theory of unoriented closed strings (from the
untwisted sector, by projecting onto the $Z_2$-invariant states) along
with unoriented open strings carrying $SO(32)$ Chan-Paton factors
(from the twisted sector). In the latter case the situation is more
subtle: first of all one should compactify to 9 dimensions on a
circle, then take the $Z_2$ quotient of the circle simultaneously with
orientation reversal of the world sheet. The result is again a theory
of unoriented closed strings in the untwisted sector, but the twisted
sector states are open strings with Dirichlet boundary conditions
which constrain their ends to lie on 8-branes in space, transverse to
the two fixed points of the 10th direction. These open strings carry
$SO(16)$ Chan-Paton factors and produce a gauge group $SO(16)\times
SO(16)$, one factor from each fixed point.

The two orientifold theories described above become equivalent when
viewed in 9 dimensions, after turning on a Wilson line in the former.
The moduli spaces are identical although Wilson lines in the former
theory correspond to shifting the end points of the open strings in
the latter.

The above picture emerges after explicit calculation of amplitudes
which have a simultaneous interpretation as closed-string tree
diagrams and open-string one-loop diagrams. Comparing the two
calculations allows one to deduce the presence of a degeneracy which
is then interpreted in terms of Chan-Paton factors. Since analogous
calculations are inaccessible in M-theory, one may wonder how to study
the twisted sector in an orientifold of it (not to mention that one
doesn't know how to define the analogue of orientation-reversal, to
start with).

The key to obtaining a consistent picture comes from two important
facts. The field content of massless modes of M-theory is in
correspondence with that of type IIA string theory in 10 dimensions,
so that one can define an orientifold by imitating its known action on
the spacetime fields and spacetime coordinates of the IIA
theory. Second, the nature of the twisted sectors can be inferred by
requiring the cancellation of gravitational anomalies in 10 dimensions
and taking into account the multiplet structure of the residual
supersymmetry. Thus one starts with 11-dimensional supergravity,
having spacetime bosonic fields $g_{MN}$ and $A_{MNP}$, compactifies
the 11th direction $X^{11}$ on a circle of radius $R_{11}$, and
considers the $Z_2$ action
\eqn\ztwo{
\eqalign{
A&\to -A\cr
X^{11} &\to - X^{11}\cr}}
The $Z_2$ invariant modes in 10 dimensions are $g_{\mu\nu}, g_{11,11}$
and $A_{\mu\nu,11}$, or in other words a metric, scalar and 2-form in
10d. This is precisely the bosonic content of the (chiral) $N=1$
supergravity multiplet in 10d. Since this multiplet suffers from a
gravitational anomaly, one expects twisted-sector states to arise
precisely in the right combination to cancel it. Such states must lie
in chiral $N=1$ gauge multiplets, and there must be two copies (one at
each fixed point), which uniquely determines the gauge group to be
$E_8\times E_8$. Many checks of this picture were made in Ref.\horwit\
and they all tend to confirm it.

This does not tell us how the gauge groups actually arise in M-theory,
though we will speculate on that later. In the following, we first
investigate other situations where something can be said about
orbifolds of M-theory.

\newsec{$T^5/Z_2$ Orbifold of M-theory, and a puzzle}
Besides the example above, gravitational anomalies can only help us if
we compactify M-theory to 6 or 2 spacetime dimensions. Accordingly we
first investigate the orientifold down to 6 dimensions on
$T^5/Z_2$. As for the $S^1/Z_2$ case, the 3-form changes sign under
the reversal of the five coordinates $X^7,X^8,X^9,X^{10},X^{11}$. The
result is the following set of massless $Z_2$-invariant fields:
\eqn\tfive{
\eqalign{
g_{MN} &\to g_{\mu\nu},~ g_{ij}=15\phi\cr
A_{MNP} &\to A_{\mu\nu i}=5 B_{\mu\nu},~A_{ijk} = 10\phi\cr}}
Thus the total bosonic spectrum in the untwisted sector is a metric, 5
2-forms and 25 scalars. Remarkably, this is enough to tell us that
we are dealing with the {\it chiral} $N=4$ supersymmetry in 6
dimensions, since that is the only one for which the above spectrum
can be decomposed into multiplets! Of course, this is consistent with
the expectation that the orientifold actually reverses the sign of
half the components of the 11-dimensional gravitino, leaving four
chiral symplectic Majorana-Weyl (SMW) gravitinoes and 20 SMW fermions
of the opposite chirality. (Note that this supersymmetry is $N=4$ in
SMW units and leads to $N=4$ supersymmetry on toroidal
compactification to four dimensions. In the literature it is equally
often called $N=2$, counted in units of Weyl spinors.)

Indeed, the spectrum obtained above decomposes into a chiral
supergravity multiplet $R(G) = (g_{\mu\nu}, B^{(ij)(-)}_{\mu\nu})$ (we
list only the bosons, the $(-)$ superscript indicates an anti-self-dual
two-form, and $(ij)$ is a composite index representing the 5 of
$USp(4)$) and five matter multiplets $R(M) = (B_{\mu\nu}^{(+)},
\phi^{(ij)})$. We now have a situation very analogous to that in 10d,
since the above structure suffers from a gravitational anomaly, and
there is a unique way to eliminate the anomaly. This corresponds to
adding 16 more matter multiplets $R(M)$. The resulting spectrum is
that of the type IIB supergravity compactified on K3\ref\towne{\TOWNE},
whose low-energy multiplet structure is the unique one consistent with
chiral $N=4$ supersymmetry and anomaly cancellation.

We would therefore conclude that the ``twisted sector'' of M-theory on
$T^5/Z_2$ provides precisely 16 matter multiplets $R(M)$, and that the
resulting theory is equivalent to the type IIB string on K3.
Before going on to test this hypothesis, however, we are faced with an
immediate puzzle. The torus $T^5/Z_2$ has 32 fixed points. Taking a
regular torus with all sides equal, one would expect the anomaly to be
distributed equally among the fixed points (and perhaps described by
open membranes with Dirichlet boundary conditions which attach their
boundaries to the fixed points). But even with our limited
understanding of M-theory, it seems impossible to obtain 16 multiplets
equally distributed among 32 fixed points, at least without
violating manifest supersymmetry and Lorentz invariance in the
formalism.

To investigate this point in more detail, consider the further
compactification to 5 dimensions on $S^1$. On one side, we have
M-theory on $T^5/Z_2\times S^1$, while on the other we expect type IIB
on $K3\times S^1$. The latter theory (for generic K3) gives a
massless spectrum containing the $N=4$ supergravity multiplet in 5d,
along with 21 copies of the $N=4$ Maxwell multiplet. For the former,
we can reverse the order of compactification. M-theory on $S^1$ gives
the type IIA string in 10d, and we should now make a $T^5/Z_2$
orientifold of that to five dimensions and compare.

This time we know what to do, since orientifolds of the type IIA
string are well-defined -- indeed, they are precisely the ones which
inspired the similar construction in 11d that we are attempting to
pursue. It has been argued in Ref.\horava\ that the orientifold of the
type $IIA/B$ string on $T^n/Z_2$ (IIA for odd $n$ and IIB for even
$n$) gives a theory, called type I$'$, whose twisted sector contains
open strings with the Chan-Paton factors of $SO(32.2^{-n})$. Since
there are $2^n$ fixed points, the full Chan-Paton group is
$\big(SO(32.2^{-n})\big)^{2^n}$. The rank of this group is formally
$\half.32.2^{-n}.2^n=16$ independent of $n$, but actually this is
valid only for $n\le6$. For $n=5$ it appears that the Chan-Paton group
has become $\big(SO(1)\big)^{32}$, and $SO(1)$ strings have no
massless states (they do have massive ones, at odd
levels\ref\schwhopk{\SCHWHOPK}). So from this discussion it appears
that the twisted sector of the type IIA string on $T^5/Z_2$
contributes no massless multiplets. Then in comparing the massless
spectrum to type IIB on $K3\times S^1$, we would find it is missing
precisely 16 Maxwell multiplets. This appears to violate the proposed
equivalence of M-theory on $T^5/Z_2$ to IIB on K3.

This picture cannot be quite right, for the following reason. The
orientifold on $T^5/Z_2$ of type IIA is, more or less by definition,
the T-dual of the type I open string on $T^5$, when all the 5
directions are dualized. The latter theory can certainly have rank-16
gauge groups (Abelian or non-Abelian) in 5d, so it must be possible
for the former to have them as well. The only thing that is impossible
in 5d or less is to have symmetrically distributed Chan-Paton factors
over 32 or more fixed points in the type I$'$ description. Indeed, the
analysis of Chan-Paton factors on orientifolds in
Refs.\ref\polcomb{\POLCOMB} and \ref\polwit{\POLWIT} implies that type
I$'$ strings always have $U(1)^{16}$ Chan-Paton factors, which can be
distributed in various ways on the orbifold, and give rise to unitary
enhanced symmetries whenever they merge, and even larger orthogonal
groups when they merge at fixed points.

Nevertheless, the question arises why M-theory produces only 16 matter
multiplets in 6d, the unique number allowed by gravitational anomaly
cancellation, despite the presence of 32 fixed points. We will now
propose a resolution to the puzzle.

\newsec{Couplings and Radii}
Let us assume despite the apparent puzzle found above, that M theory
on $T^5/Z_2$ is equivalent to the type IIB string on K3. Since M
theory has no moduli or couplings, all the parameters of the resulting
6d theory will come from the toroidal orbifold. On the IIB side, there
is a moduli space labelled by $O(21,5,R)/O(21,R)\times
O(5,R)$\ref\seiberg{\SEIBERG}\ref\aspmor{\ASPMOR}. The duality group
acting on this is $O(21,5,\ZZ)$, and this includes, among other
things, strong-weak coupling duality transformations on the coupling
constant in 6 dimensions.

We will find a relation between the M-theory radii and this coupling
constant, and check it in two different ways. This will enable us to
propose a resolution to the puzzle encountered in the previous
section. 

M-theory compactified on a circle of radius $R_{11}$ is believed to be
equivalent to type IIA string theory in 10d\towna\wittena\ 
with the string coupling $\lambda_{10}$ given by\wittena:
\eqn\iiaten{
\lambda_{10}^{2/3} = R_{11}}
If we further compactify the IIA theory on $T^4$ to 6 dimensions,
the coupling constant in 6d is given by the usual relation
\eqn\usual{
\lambda_6 = {\lambda_{10}\over \sqrt{vol(T^4)}}}
Assume a rectangular torus. The compactification radii $R_i,
i=7,\ldots,10$ as measured in the M-theory metric are related to the
radii $R^{(10)}_i$ measured in the string metric of the IIA theory
by:
\eqn\radii{
R_i = {R^{(10)}_i\over \lambda_{10}^{1/3}}}
as follows from the Weyl scaling between the two metrics. The volume
of $T^4$ in Eq.\usual\ above is the one measured in the string metric,
so we find
\eqn\coupling{
\lambda_6 = {\lambda_{10}^{1/3}\over (R_7 R_8 R_9 R_{10})^{1/2}} =
\left({R_{11}\over R_7 R_8 R_9 R_{10}}\right)^{1/2} }

Now in this discussion we have been working with a torus -- and not
orbifold -- compactification of M-theory. But relations between
couplings and radii such as the ones derived above are insensitive to
orbifolding, so we can conclude that the above relations hold also for
M-theory on $T^5/Z_2$ and hence that $\lambda_6$ is the coupling of
the type IIB theory on K3. The asymmetry in the five radii
of the torus-orbifold will have a natural interpretation in terms of
the type IIB theory, as we will see below.

One way to check the above relation, and our proposal for the M-theory
orbifold in general, is the following. If the M-theory on $T^5/Z_2$ is
indeed equivalent to type IIB on K3, then the equivalence must be
present on further compactification of both sides on $S^1$, as
mentioned before. The type IIB theory on $K3\times S^1$ has been
argued\wittena\ to be equivalent to the heterotic string toroidally
compactified to 5 dimensions, with the identification
\eqn\twobhet{
R_{h,6} = {1\over\lambda_{B,6}}.}
where the $h$ and $B$ subscripts refer to the heterotic and type IIB
strings respectively. 

On the other side, we get a type I$'$ theory of open and closed
strings. Relationships between toroidally compactified type I, type I$'$
and heterotic strings have been recently explored\polwit. They are all
equivalent, with the maps being T-duality from type I$'$ to type I, and
strong-weak coupling duality from type I to
heterotic\ref\dabholkar{\DABHOLKAR}\ref\hull{\HULL}. One can derive
relationships between the radii and coupling constants of type I$'$ and
heterotic strings using type I as an intermediate step. For $n$
compactified dimensions, the result is (see Eq.(21) of Ref.\polwit)
\eqn\iprimehet{
\left(R_{I',i}\right)^{n-2} = \lambda_{I',10} {\prod_{i=11-n}^{10}
R_{h,i}\over \left(R_{h,i}\right)^{n-2}}}
This is easily inverted to express each heterotic string radius in
terms of all the type I$'$ radii and the 10d coupling:
\eqn\hetiprime{
R_{h,i} = \left({\prod_{i=11-n}^{10} R_{I',i}\over 
\lambda_{I',10}}\right)^{1/2} {1\over R_{I',i}}}
As before, the radii $R_{I',i}$ are related to those measured in
the M-theory by $R_{I',i} = \lambda_{I',10}^{1/3} R_i$. Inserting
this, we find in particular that
\eqn\rhet{
R_{h,6} = \left({\prod_{i=11-n}^{10} R_i\over R_6}\right)^{1/2}
\lambda_{I',10}^{(n-5)/6}}
For $n=5$, the case at hand, the coupling constant dependence neatly
drops out. Finally, we recall that what we are calling $R_6$ here is
one of the radii of the 5-torus orbifold, which has been exchanged
with $R_{11}$ when we interchanged the orders of compactification on
$T^5/Z_2$ and $S^1$. Thus we should now rename $R_6$ as
$R_{11}$. Combining Eqs.\rhet\ and \twobhet\ then leads to
\eqn\confirm{
\lambda_{B,6} = \left({R_{11}\over 
R_7 R_8 R_9 R_{10}}\right)^{1/2} }
in perfect agreement with Eq.\coupling\ above.

The asymmetry in the above relation, alluded to earlier, leads one to
suspect that the 5-torus-orbifold will be generically anisotropic.
This seems to be an important clue for the resolution of our puzzle of
the previous section. To clarify it further, and to provide yet
another consistency check on Eq.\coupling, we compare a suitable
sector of the M-theory action compactified on $T^5$ and projected to
$Z_2$ invariant states, with the relevant terms in the action of type
IIB on K3. Start with the bosonic sector of the 11d supergravity
(neglecting the Chern-Simons term, since after $Z_2$ projection this
will only contribute coupling terms to scalars which we are ignoring):
\eqn\maction{
S_M^{(11)} = \int~d^{11}x \sqrt{G^{(11)}}\left(R + |dA^{(3)}|^2\right) }
and choose the 11d metric to be
\eqn\elevmetric{
ds^2 = G_{\mu\nu}^{(6)} dx^\mu dx^\nu + \sum_{i=7}^{11}
e^{2\gamma_i} (dx^i)^2
+ \ldots}
Here we have restricted ourselves to a rectangular 5-torus, so the
angular terms do not appear. We also neglect the Maxwell field coming
from the off-diagonal graviton, since this will be projected out by
$Z_2$. The action in 6d is then
\eqn\msixd{
S_M^{(6)} = \int~d^6 x \sqrt{G^{(6)}}e^{\sum_i \gamma_i} 
\left(R + \sum_{i=7}^{11} e^{-2\gamma_i}|dB^i|^2\right) + \ldots }
where $B_{\mu\nu}^i = A^{(3)}_{\mu\nu i}$ are the five 2-form fields
in 6d. 

To compare with type IIB on K3, recall that in this case there are
altogether 21 anti-self-dual and 5 self-dual $B$-fields, of which we
set 16 of the anti-self-dual fields to zero since on the M-theory side
they are expected to come from the twisted sector, over which we have
no control. Combining the remaining 10 pairwise, we have 5
unconstrained $B$-fields left. The important point is that one of
these comes from the NS-NS sector while the other 4 (as well as the 16
anti-self-dual ones that we neglected) are Ramond-Ramond.

There are also 25 scalars in the collection of multiplets that we are
considering, but we set all of them to zero except the dilaton. Then on
general grounds, the relevant part of the IIB action in 6d will be
\eqn\twobsixd{
S_{IIB}^{(6)} = \int~d^6 x \sqrt{g}\left[ e^{-2\phi} 
\left(R +  |dB^{11}|^2\right) +  \sum_{i=7}^{10} |dB^i|^2\right] + \ldots }
where $B^{11}$ has been chosen to be the unique NS-NS 2-form and the
other $B^i$ are R-R.

If we now compare the two 6d actions above, we can hope to express all
the five compactification radii as functions of the dilaton, hence of
the IIB string coupling. This relation will be valid only in
the region of moduli space where all the other scalars have been set
to zero, but it should serve as a qualitative guide to the
relationship between the two theories. Moreover, whatever relationship
emerges should of course satisfy Eq.\coupling\ above.

Following a by now well-established procedure, we make a Weyl
rescaling to relate the metric $G^{(6)}$ coming from M-theory and the
metric $g$ of type IIB on K3:
\eqn\weyl{
G^{(6)}_{\mu\nu} = e^{-\gamma} g_{\mu\nu} }
where $\gamma$ is to be determined. Matching the two sides leads to
the set of equations
\eqn\match{
\eqalign{
-2\gamma + \sum_i \gamma_i &= -2\phi\cr
\sum_i \gamma_i -2\gamma_{11} &= -2\phi\cr
\sum_i \gamma_i - 2 \gamma_j &= 0,~~j=7,\ldots,10\cr}}
whose solution is
\eqn\soln{
\eqalign{\gamma_{11} &= {2\over3}\phi\cr
\gamma_i &= -{1\over 3}\phi,~~i=7,\ldots,10 \cr}}

From the M-theory point of view the compactification radii are just
$R_i=e^{\gamma_i}$. Thus we have found that 
\eqn\compradii{
\eqalign{
R_{11} &= \lambda_6^{2/3}\cr
R_i &= \lambda_6^{-1/3},~~i=7,\ldots,10\cr}}
with $\lambda_6=e^\phi$. It is reassuring that this set of values
satisfies Eq.\coupling, as one can easily check.

At this point we see that (at least for the region of moduli space on
which we are concentrating) the torus-orbifold scales anisotropically
with $\lambda_6$. For weak coupling, four of the sides are very large
and one is very small. Moreover,
\eqn\ratio{
{R_{11}\over R_i} = \lambda_B,~~i=7,\ldots,10}
so the ratio of the short side to the long ones is proportional to the
string coupling. The conclusion that we draw from this is that for
weak coupling, the torus is highly ``squashed'', and approximates a
four-dimensional, rather than five-dimensional, manifold in this
limit. 

We can now propose, as a resolution to our puzzle of the previous
section, that paradoxes based on equal distribution of Chan-Paton
factors or their M-theory equivalents, among the 32 fixed points of
$T^5/Z_2$ are not valid, since a regular 5-torus is highly
non-generic. Indeed, it is precisely in the region where the 6d
coupling is of order 1 (the ``self-dual'' point of the IIB string)
where the torus regains isotropy, but this region is out of reach of
perturbation theory (unlike both the strong and weak coupling regions,
which can be exchanged by the S-duality contained in $O(5,21,\ZZ)$).

As the torus-orbifold degenerates, the 32 fixed points degenerate to
16 pairs with relative spacing of order the string coupling within
each pair. So one may hope that M-theory on this space produces a
multiple of 16, and not necessarily 32, copies of the twisted-sector
multiplet. Thus it is consistent to assume that 16 chiral $N=4$
multiplets in 6d are produced by M-theory in the twisted sector, as
required to cancel gravitational anomalies.

The same phenomenon applies to the squashed $T^5/Z_2$ orientifold of
the type IIA string to 5 dimensions. The computation of Ref.\horava\
for this case should lead to the conclusion that the $(U(1)^{16}$
Chan-Paton group arises symmetrically since there are now only 16 and
not 32 fixed points. (It may be worth studying this point more
closely, though we will not do so here.) Thus the 5d theory will have
16 extra $N=4$ Maxwell multiplets in 5d, making a total of 21 (since 5
come from the untwisted sector). This is precisely what one gets from
the type IIB theory on $K3\times S^1$ at generic points of the K3
moduli space.

If we turn on the other moduli, we can imagine more general
degenerations of the squashed torus. For example, two or more sides
could degenerate together. In these cases we will have $2^{5-k}$ fixed
points where $k=2,3,4,5$ is the number of sides degenerating. Then of
the 16 ``missing multiplets'', $2^{k-1}$ will come from each fixed
point. {}From general principles of orientifolding, one expects that
type IIA on such an orientifold should have enhanced gauge symmetries
coming from Chan-Paton factors. Going back to M-theory, this provides
an indirect argument that such squashed tori with $k\ge 2$ correspond
to special points in the K3 moduli space, namely the points where IIA
would acquire enhanced gauge symmetry or IIB would acquire a
non-perturbative self-dual string of vanishing tension. Thus it is
reasonable to believe that the ``generic'' $T^5/Z_2$ appropriate for
compactification of M-theory to 6 dimensions is the one with a single
squashed dimension.

It is noteworthy that anisotropic tori have recently played an
important role\polwit\ in understanding some apparent paradoxes in the
proposed strong-weak coupling duality between heterotic and type I
strings. 

\newsec{Other Orbifolds of M-theory}
In 10 dimensions, orientifolds on $T^n/Z_2$ makes sense in type IIA
for odd $n$ and type IIB for even $n$. M-theory is of course more
closely related to type IIA than to type IIB. In particular, the
$S^1/Z_2$ and $T^5/Z_2$ orbifolds of type IIA theory and M-theory
have been discussed and compared in the preceding sections, and it is
evident that there are both intriguing similarities and notable
differences between the string and M-theory cases. Perhaps the most
striking difference is that in the type IIA case one gets non-chiral
theories (hardly surprising since one also lands in odd dimensions)
but in M-theory the two known orbifolds lead to chiral string
theories. 

Here we investigate the remaining $T^n/Z_2$ orbifolds of M-theory, but
will come to the disappointing conclusion that most of them break
supersymmetry completely. 

The first potentially interesting example is $T^3/Z_2$.  In this case
one might expect to find an interesting theory (although there will be
no help from gravitational anomalies). However, the massless spectrum
of bosonic fields in the untwisted sector is easily found to be
$(g_{\mu\nu}, 3B_{\mu\nu}, 7\phi)$. This does not break up into
supermultiplets, since $N=1$ supersymmetry in 8 dimensions admits only
a single 2-form field in the supergravity multiplet and none in the
matter multiplet. Thus this orbifold breaks supersymmetry
completely. This may appear surprising, since there must be a
$T^3/Z_2$ orbifold of the type IIA string in 10d which preserves
half the supersymmetries, namely the T-dual of the type I string on
$T^3$ when all 3 directions are dualized. This theory would then be
expected to agree with M-theory on $T^3/Z_2\times S^1$. 

The way this paradox is avoided is that for the $T^3/Z_2$ orbifold of
type IIA, the $Z_2$ action on the spacetime fields is
\eqn\action{
\eqalign{
B &\to -B\cr
A &\to -A\cr
A^{(3)} &\to A^{(3)}\cr}}
(with $B,A$ and $A^{(3)}$ being the 2-form, 1-form and 3-form of the
theory), unlike the orbifolds on $S^1/Z_2$ and $T^5/Z_2$ where the
action is
\eqn\otheraction{
\eqalign{
B &\to -B\cr
A &\to A\cr
A^{(3)} &\to -A^{(3)}\cr}}
The latter $Z_2$ action ``lifts'' to M-theory since $B$ and $A^{(3)}$
transform similarly, and they merge into a single 3-form field in 11d.
However, the former action does not, since $B$ and $A$ have different
M-theoretic origins.

The $T^7/Z_2$ orbifold of M-theory similarly breaks supersymmetry
completely. One heuristic way to think of this phenomenon is that for
$T^5/Z_2$, the part of the transformation living in the first 10
dimensions is exactly the orbifold limit of K3 and has $SU(2)$
holonomy, while for $S^1/Z_2$ there is no such part and hence no
holonomy. For the $T^3/Z_2$ and $T^7/Z_2$ cases, the part of the
transformation visible below 10d corresponds to minus the identity
matrix in one and three complex dimensions respectively, hence they do
not have the right holonomies ($SU(1)$ and $SU(3)$ respectively) to
preserve supersymmetry.

For $T^9/Z_2$ the situation is more interesting. From M-theory, one
finds that the $Z_2$-invariant spectrum in 2d contains 129 scalars.
This may seem puzzling, but it comes from the fact that the metric in
2d, after removing its trace part, has formally $-1$ degree of
freedom. So one of the 129 scalars should be combined with the metric,
giving a supergravity sector with no on-shell degrees of freedom.
Similarly, the 16 left-moving Majorana-Weyl gravitinos have formally
$-16$ degrees of freedom, and combine with 16 right-moving
Majorana-Weyl fermions to fit into the supergravity multiplet. The
total gravitational anomaly\ref\alvwit{\ALVWIT} of this multiplet is
proportional to $16.{23\over 24} + 16.{1\over 24}$ which equals 16.

This leaves behind 128 scalars, which assemble into 16 real scalar
multiplets of chiral $N=8$ supersymmetry: $(\phi^i, \lambda^i_R)$ with
$i=1,\ldots,8$. The anomaly of each of these multiplets is
proportional to $8.{1\over 24}$ so altogether one has an anomaly of
${16\over 3}$. Combining the supergravity and matter multiplets, the
total gravitational anomaly is proportional to $16 + {16\over
3}={64\over 3}$. This can be cancelled by 512 left-moving spin-$\half$
fermions. Since there are now $2^9=512$ fixed points of the orbifold,
it is conceivable that one left-moving fermion can come from each of
the fixed points. These fermions would be singlets of the right-moving
$(0,8)$ supersymmetry\foot{An observation in an earlier version of our
paper, that these fermions should lie in supersymmetry multiplets, has
been corrected.}.

One piece of evidence to support this hypothesis comes from the
expression for the coupling constant of the 2d theory. Repeating the
arguments given before but now for compactification down to an
arbitrary number of dimensions $d$, one finds that the coupling
constant in $d$ dimensions is given by 
\eqn\arbit{ \eqalign{\lambda_d 
&= {\lambda_{10}\over \sqrt{vol(T^{10-d})}}\cr
&={\lambda_{10}^{1-(10-d)/6}\over \sqrt{R_{d+1}\ldots R_{10}}}\cr
&={R_{11}^{(d/4)-1}\over \sqrt{R_{d+1}\ldots R_{10}}}\cr}} 
For the special case $d=2$, one finds 
\eqn\specialcase{ \lambda_2 = {1\over
\sqrt{R_3 \ldots R_{11}}}} 
and remarkably, in this case the formula is totally symmetric in the 9
radii! This suggests that this is the unique situation where the torus
is isotropic, and the twisted sector contribution can be expected to
arise symmetrically from all the 512 fixed points. One may ask whether
the resulting theory corresponds to a known compactification of the
type IIB string down to 2 dimensions. The obvious choices $K3\times
K3$ and the Joyce 8-manifold of $Spin_7$ holonomy are ruled out since
they preserve $1/4$ and $1/16$ of the supersymmetries respectively,
while we want to preserve exactly half. Perhaps the right manifold is
something like ${K3\times K3\over Z_2}$, but we will not pursue this
point further here\foot{Actually, it is easy to check that the type
IIB string on the orbifold $T^8/Z_2$ gives the right spectrum of
states.}.

The even-dimensional orbifolds of M-theory on $T^n/Z_2$ for
$n=2,4,6,8$ do not correspond to orientifolds, but to the analogues of
conventional orbifolds in string theory. The action is invariant under
change of sign of an even number of directions without changing the
sign of the 3-form field. So we will not discover analogues of
Chan-Paton factors in this case. Moreover, these orbifolds all land us
in odd dimensions, where anomaly considerations are not useful. 

In any case, for $n=2$ and $n=6$, supersymmetry is broken, while for
$T^4/Z_2$ one finds that the untwisted sector decomposes into an $N=2$
supergravity multiplet and 3 $N=2$ Maxwell multiplets. Comparing to
the spectrum of M-theory on K3, we find precisely 16 missing Maxwell
multiplets. So, {\it if} M-theory on $T^4/Z_2$ is to be the same as
M-theory on K3 (in the appropriate orbifold limit of K3), then 16
Maxwell multiplets must come from the twisted sector. But these extra
multiplets, if they appear, are not dictated by something as
compelling as anomaly cancellation.

\newsec{Conclusions}
We have seen that one can define ``orientifolds'' of M-theory on
$T^n/Z_2$ leading to supersymmetric theories for $n=1,5$ and 9 (the
first of these cases was discovered in Ref.\horwit). All lead to
chiral theories and one can predict the twisted-sector states from
considerations of cancellation of gravitational anomalies. 

The $n=5$ case corresponds to type IIB on K3, though the
compactification torus is generically anisotropic and virtually
four-dimensional. This case should be worth studying in more detail
than we have done here, particularly in order to understand the full
correspondence between the moduli spaces for the M-theory and IIB
compactifications. It is tantalizing that of the $SO(21,5)$ group
occurring in the latter theory, the compactification torus for
M-theory can only provide moduli for an $SO(5,5)$ part, while the
twisted sector should provide another 16 degrees of freedom to enhance
$SO(5,5)$ to $SO(21,5)$. This is reminiscent of Narain's
observation\ref\narain{\NARAIN}\ that for the heterotic string, the 16
dimensional internal-symmetry lattice in the heterotic string can mix
nontrivially with the $(n,n)$ spacetime lattice after compactification
on an $n$-dimensional torus, to yield a $(16+n,n)$ lattice.

The interpretation of our results in terms of membranes of M-theory
wrapping around the orbifolds, along the lines of Ref.\horwit, remains
to be worked out.

Some of the interesting points raised by our analysis
are the following. First of all, M-theory seems to easily yield {\it
chiral} compactifications, reversing years of prejudice against
11-dimensional supergravity for its inability to provide chiral
theories in lower dimensions. (This fact is also evident, although in
a more subtle manner, from the fact that M-theory on a 2-torus gives
the 10d type IIB string in a suitable limit\aspina\schwarza). So it
should be worthwhile to investigate more sophisticated orbifolds than
the ones we have considered (quotienting by groups larger than $Z_2$),
particularly 7-dimensional ones, and see how this fits with the known
Calabi-Yau compactifications of 10d string theory. Clearly this is
much more difficult because of both the absence of gravitational
anomalies in four dimensions and the reduced supersymmetry ($N=1$).

The other intriguing question is what do we learn about the Chan-Paton
factors, or rather their analogues, for M-theory. Evidently, whatever
these are, they can produce $E_8\times E_8$ gauge groups, and also
6-dimensional supermultiplets containing anti-self-dual 2-forms, both
possibilities being out of reach of ordinary open strings. One line of
thought might be to consider open membranes with free fermions living
on their boundary, analogous to the interpretation of Chan-Paton
factors for open strings in terms of free fermions living on the
string boundary\ref\marcsag{\MARCSAG}. The worldvolume boundary field
theory of the membrane would then be a two-dimensional conformal field
theory of free fermions, and we know that with suitable spin
structures this can indeed produce $E_8$ gauge groups, unlike the
one-dimensional boundary theory of free fermions arising in open
string theory. Whether this can be made more precise, and whether the
same construction can yield anti-self-dual 2-forms in six dimensions,
remains to be investigated.
\bigskip
\noindent{\bf Note Added in Proof}

In a preprint that appeared on hep-th soon after this one,
E. Witten\ref\witnew{\WITNEW} has independently made many of the
observations in this paper, and also proposed an interesting picture
of how the twisted sector states could arise from five-branes. After
reading Ref.\witnew\ we have made a small correction in Section 4
above, see the first footnote there.
\bigskip
\noindent{\bf Acknowledgements} 

We are grateful to Camillo Imbimbo and K.S. Narain for useful
discussions.

\listrefs   
\bye